
\magnification\magstep1
\newdimen\itemindent \itemindent=32pt
\def\textindent#1{\parindent=\itemindent\let\par=\resetpar%
\indent\llap{#1\enspace}\ignorespaces}

\let\oldpar=\par
\def\resetpar{\oldpar\parindent=20pt\let\par=\oldpar}

\font\ninerm=cmr9 \font\ninesy=cmsy9
\font\eightrm=cmr8 \font\sixrm=cmr6
\font\eighti=cmmi8 \font\sixi=cmmi6
\font\eightsy=cmsy8 \font\sixsy=cmsy6
\font\eightbf=cmbx8 \font\sixbf=cmbx6
\font\eightit=cmti8
\def\eightpoint{\def\rm{\fam0\eightrm}
  \textfont0=\eightrm \scriptfont0=\sixrm \scriptscriptfont0=\fiverm
  \textfont1=\eighti  \scriptfont1=\sixi  \scriptscriptfont1=\fivei
  \textfont2=\eightsy \scriptfont2=\sixsy \scriptscriptfont2=\fivesy
  \textfont3=\tenex   \scriptfont3=\tenex \scriptscriptfont3=\tenex
  \textfont\itfam=\eightit  \def\it{\fam\itfam\eightit}%
  \textfont\bffam=\eightbf  \scriptfont\bffam=\sixbf
  \scriptscriptfont\bffam=\fivebf  \def\bf{\fam\bffam\eightbf}%
  \normalbaselineskip=9pt
  \setbox\strutbox=\hbox{\vrule height7pt depth2pt width0pt}%
  \let\big=\eightbig  \normalbaselines\rm}
\catcode`@=11 %
\def\eightbig#1{{\hbox{$\textfont0=\ninerm\textfont2=\ninesy
  \left#1\vbox to6.5pt{}\right.\n@space$}}}
\def\vfootnote#1{\insert\footins\bgroup\eightpoint
  \interlinepenalty=\interfootnotelinepenalty
  \splittopskip=\ht\strutbox %
  \splitmaxdepth=\dp\strutbox %
  \leftskip=0pt \rightskip=0pt \spaceskip=0pt \xspaceskip=0pt
  \textindent{#1}\footstrut\futurelet\next\fo@t}
\catcode`@=12 %

\magnification\magstep1
\def\pmb#1{\setbox0=\hbox{#1}%
 \kern-.025em\copy0\kern-\wd0
 \kern.05em\copy0\kern-\wd0
 \kern-.025em\raise.0433em\box0 }
\def \bxi{\pmb{$\xi$}}
\def \olr{{\raise7pt\hbox{$\leftrightarrow  \! \! \! \! \!$}}}
\def \ollr{{\raise7pt\hbox{$\leftrightarrow  \! \! \! \! \! \!$}}}
\def \de{\delta}
\def \si{\sigma}

\def \ga{\gamma}
\def \nab{\nabla}
\def \hnab{{\hat \nabla}}

\def \pr{\partial}
\def \hx{{\hat x}}
\def \bx{{\bf x}}
\def \hbx{{\hat {\bf x}}}

\def \d2{{\textstyle {1 \over 2}} d}

\def \rO{{\rm O}}
\def \vep{\varepsilon}
\def \half{{\textstyle {1 \over 2}}}

\def \quar{{\textstyle {1 \over 4}}}

\def \hh{{1\over 2}}
\def \ts{ \textstyle}

\def \tx{{\tilde x}}

\def \R{{\cal R}}

\def \L{{\cal L}}
\def \O{{\cal O}}
\def \M{{\cal M}}

\def \hx{{\hat x}}

\def \hga{{\hat {\gamma}}}

\def \b{{\rm b}}

\input epsf
\parskip 5pt
\font \bigbf=cmbx10 scaled \magstep1
{\nopagenumbers
\rightline{DAMTP/93-1}
\vskip 2truecm
\centerline {\bigbf Energy Momentum Tensor in Conformal Field Theories}
\vskip 5pt
\centerline {\bigbf Near a Boundary}
\vskip 2.0 true cm
\centerline {D.M. McAvity and H. Osborn}
\vskip 10pt
\centerline {Department of Applied Mathematics and Theoretical Physics,}
\centerline {Silver Street, Cambridge, UK}
\vskip 2.0 true cm
The requirements of conformal invariance for the two point function of the
energy momentum tensor in the neighbourhood of a plane boundary are
investigated, restricting the conformal group to those transformations
leaving the boundary invariant. It is shown that the general solution
may contain an arbitrary function of a single conformally invariant variable
$v$, except in dimension 2. The functional dependence on $v$ is determined for
free scalar and fermion fields in arbitrary dimension $d$ and also to leading
order in the $\vep$ expansion about $d=4$ for the non Gaussian fixed point in
$\phi^4$ theory. The two point correlation function
of the energy momentum tensor and a scalar field is also shown
to have a unique expression in terms of $v$ and the overall coefficient
is determined by the operator product expansion. The energy momentum tensor on
a general curved manifold is further discussed by considering variations of
the metric. In the presence of a boundary this procedure naturally defines
extra boundary operators. By considering diffeomorphisms these are related
to components of the energy momentum tensor on the boundary. The implications
of Weyl invariance in this framework are also derived.
\vfill\eject}
\pageno=1

\leftline{\bigbf 1 Introduction}
\medskip

Besides the usual bulk critical phenomena in a statistical mechanical system
there are theoretically interesting and experimentally measurable effects
associated with the presence of a surface, or the consequences of finite size,
which are inevitably present in any realisable physical system.
There are possible phase transitions corresponding to
surface ordering such that, at or close to a critical point
where the correlation length is large and microscopic details are unimportant,
there are additional critical exponents describing the behaviour of
correlation functions of operators on or in the neighbourhood of the boundary
surface. These may also be discussed within the framework of continuum scale
invariant quantum field theories just as for conventional considerations of
bulk critical phenomena [1]. Furthermore, as Cardy has shown [2], it
is similarly
natural to impose also the consequences of conformal invariance which is
present at the critical point assuming the trace of the energy momentum
tensor vanishes. This restricts the functional form of correlation functions
in the neighbourhood of the boundary although the constraints are less than
in the bulk since, assuming a plane boundary for $d$-dimensional Euclidean
space with $d>2$, the symmetry group which leaves the boundary invariant is
reduced from the full conformal group $O(d+1,1)$ to $O(d,1)$. Hence
in general even two point functions of primary operators are not determined
uniquely up to a constant factor but may contain an arbitrary function.

Nevertheless Cardy also subsequently obtained significant relations [3] between
critical exponents and the coefficients of the universal terms in the Casimir
energy for simple geometries. In part his arguments depended on the assumption
that the functional dependence of the two point function for the energy
momentum tensor was essentially unique for conformal theories in the
neighbourhood of a boundary for general $d$. This derivation has recently been
criticised [4] since the required assumption  was incompatible with the
results of explicit  calculations to $\rO (\vep)$ at
the non Gaussian fixed point present in scalar $\phi^4$ theory when $d=4-\vep$.

In this paper we re-analyse the implications of conformal invariance for two
point functions involving the energy momentum tensor in the presence of a plane
boundary. Although the conditions we derive are necessary rather than
sufficient they allow for an arbitrary function to be present in the two
point function, except when $d=2$, and this freedom appears to be required to
accommodate the results of calculations in specific conformal field theories
for $d>2$ (even for free scalar fields the two point function depends on the
particular boundary conditions, compatible with conformal invariance, obeyed
by the scalar field).

In the next section we therefore derive the consequences of conformal
invariance for the two point function of the energy momentum tensor near
a boundary by obtaining differential equations for the dependence on a variable
$v$, formed from the spatial arguments $x,x'$ and which is an invariant under
conformal transformations leaving the boundary invariant. For either $x$ or
$x'$ on the boundary $v=1$ whereas for both $x,x'$ at large distances from
the boundary compared with their separation $v\to 0$. In this limit the two
point function may be related to that for the bulk critical system neglecting
any boundary effects. For $d=2$ the equations have a unique functional
solution but not if $d>2$. We also consider the two point function of the
energy momentum tensor and a scalar field of arbitrary dimension which may
be non zero in the presence of a boundary and for which there is a unique
functional solution for the dependence on $v$ for any $d$. In section 3 we
determine the functional dependence on $v$ for free scalar fields, with either
Dirichlet or Neumann boundary conditions which maintain conformal invariance,
corresponding to the simple Gaussian fixed point in scalar field theories
for any $d$. We also calculate the corresponding expressions to $\rO
(\vep)$  at
the non trivial fixed point in the $O(n)$ $\phi^4$ theory which is found in
the $\vep=4-d$ expansion around $d=4$. Restricting to the two point function of
$T_{nn}$, where $n$ denotes the normal component on the boundary, we are able
to recover the results of Eisenriegler {\it et al.} [4].

In section 4 we further consider general results for the energy momentum tensor
$T_{\mu \nu}$ as defined by variations with respect to the metric
$g_{\mu \nu}$ on a curved manifold with a smooth boundary when variations of
the induced metric, and also related geometric quantities  on the boundary,
are taken into account. The implications of diffeomorphism invariance, which
now go beyond the usual conservation equation $\nab^\mu T_{\mu \nu} = 0$, and
Weyl invariance under local rescalings of the metric, which extend the
traceless
condition $g^{\mu \nu} T_{\mu \nu}=0$, are derived. These conditions relate
components of the energy momentum tensor on the boundary to new local
boundary operators present in any field theory defined for arbitrary smooth
boundaries. The results are verified for a general scalar field theory treated
classically or to zeroth order in the loop expansion. For completeness a brief
summary of the essential results used here arising from a covariant geometrical
treatment for smoothly curved boundary surfaces is given in appendix
A. Appendix
B contains a discussion of the extension of the treatment in section 4 to
fermion fields while appendix C contains some of the salient details of the
$\rO (\vep)$ calculations necessary to obtain the results in section 3.
\bigskip
\leftline{\bigbf 2 Conformal Invariance with a Plane Boundary}
\medskip

For a flat $d$-dimensional space with coordinates $x_\mu = (x_1,\bx)$ we
assume a plane boundary at $x_1=0$. It is then necessary to restrict the
conformal group to the subgroup leaving $x_1=0$ invariant [2]. Besides $d-1$
dimensional translations, $O(d-1)$ rotations and scale transformations
$$
x_i \to x_i + a_i \ , \quad x_i \to R_{ij} x_j \ , \quad x_\mu \to \lambda
x_\mu \ ,
\eqno (2.1) $$
where $i=2,3, \dots$, this restriction also allows special conformal
transformations
$$
x_\mu \to {x_\mu + b_\mu x^2\over \Omega (x)} \ , \quad
\Omega (x) = 1 + 2b{\cdot x} + b^2 x^2 \ ,
\eqno (2.2) $$
so long as $b_1 =0$. In this case it is easy to see that for two points
$x_\mu,x'_\mu$
$$
(x-x')^2 \to {(x-x')^2 \over \Omega(x) \Omega(x')} \ , \quad
x_1 \to {x_1 \over \Omega(x) } \ , \quad
x'_1 \to {x'_1 \over \Omega(x') } \ ,
\eqno (2.3) $$
so that it is straightforward to form an invariant under the restricted
conformal group which we take as
$$
v^2 = {(x-x')^2 \over (\bx - \bx')^2 + (x_1 + x'_1)^2} \ .
\eqno (2.4) $$
In consequence for a scalar operator $\O(x)$ of dimension $\eta$ the
connected two point function has the general form, as required by
conformal invariance near a boundary,
$$
\langle \O(x) \O(x')\rangle = {1\over (x-x')^{2\eta}}\, F(v) \ ,
\eqno (2.5) $$
so that for $x_1 =0$ or $x'_1=0$ the magnitude is given by $F(1)$ (if $\O(x)$
obeys Dirichlet boundary conditions $F(1)=0$) whereas the scale of the bulk
amplitude neglecting surface effects is given by $F(0)$.

For the energy momentum tensor $T_{\mu \nu}$ under a general conformal
transformation $\tx \to x$
$$
T_{\mu \nu} (\tx) \to \Omega(\tx)^d \R_{\mu \alpha}(\tx) \R_{\nu \beta}(\tx)
T_{\alpha \beta} (\tx) \ , \quad \R_{\mu \alpha}(\tx) = \Omega (\tx)
{\pr x_\mu \over \pr \tx_\alpha} \ , \quad \R_{\mu \alpha} \R_{\nu \alpha}
= \delta_{\mu \nu} \ .
\eqno (2.6) $$
It is convenient to write the two point function for the energy momentum
tensor in the form (in this semi-infinite geometry $\langle T_{\mu \nu}
\rangle =0$)
$$
\langle T_{\mu \nu}(x) T_{\si \rho}(x') \rangle = {1\over (x-x')^{2d}} \,
A_{\mu \nu \si \rho} ({\bf s},x_1, x'_1) \ , \quad {\bf s} = \bx -
\bx' \ .
\eqno (2.7) $$

\vskip 0.1in \hskip 1.0in
\epsfbox{fig1.eps}

By a translation and
an appropriate special conformal transformation of the form (2.2), with
$b_1 =0$, we may choose $\bx=\bx'={\bf 0}$ so that $x,x'$ lie on a
perpendicular to the boundary as in fig. 1. In this case, by invariance
under rotations and scale transformations as in (2.1) which preserve this
perpendicular geometry, we may write
$$ \eqalign{
& A_{1111} = \alpha \ , \quad A_{ij11} = \beta \, \de_{ij} \ ,
\quad A_{11k\ell} = \beta' \, \de_{k\ell} \ ,
\quad A_{i1k1} = \gamma \, \de_{ik} \ , \cr
& A_{ijk\ell} = \de \, \de_{ij} \de_{k\ell} + \epsilon (\de_{ik} \de_{j\ell}
+ \de_{i\ell} \de_{jk} ) \ , \cr}
\eqno (2.8) $$
with other components given by symmetry ($A_{\mu \nu \si \rho} =
A_{\nu \mu \si \rho} = A_{\mu \nu \rho \si}$). The coefficients
$\alpha,\beta,\beta',\gamma,\de,\epsilon$ are functions of
$$
v= {y-y'\over y+y'} \ , \quad x_1 = y > x'_1 = y' \ .
\eqno (2.9) $$
Assuming the form (2.8) it is trivial to obtain the conditions for
tracelessness
$$\eqalignno{
&\alpha + (d-1) \beta = 0 \ , \quad \beta = \beta' \ , & (2.10a) \cr
& \beta + (d-1) \de + 2 \epsilon = 0 \ . & (2.10b) \cr}
$$

In addition it is necessary to impose the conservation
equation $\pr_\mu \langle T_{\mu \nu} T_{\si \rho}\rangle = 0$.
To derive these extra conditions\footnote{*}{The following arguments
are an adaptation of a similar discussion of conformal invariance
requirements on three point functions for general $d$ [5].}
it is necessary to extend the form (2.8),
valid for ${\bf s} = {\bf 0}$, to ${\bf s}\ne {\bf 0}$. For an infinitesimal
transformation leaving $x'_\mu = (y' , {\bf 0})$ invariant
$$ \eqalign {
& \de x_i = b_i x^2 - 2 x_i {\bf b} {\cdot \bx} - b_i y'^2 \ , \quad
\de x_1 = - 2 x_1  {\bf b} {\cdot \bx} \ , \cr
& \de (y', {\bf 0}) = 0 \ , \quad \de (y, {\bf 0}) = ( 0 , {\bxi})
\ , \quad {\bxi} = {\bf b} ( y^2 - y'^2) \ , \cr}
\eqno (2.11) $$
then the $O(d)$ transformation matrices as defined by (2.6) at $x$ and $x'$
respectively are
$$
\R_{\mu \alpha} = \pmatrix { 1 & - 2b_j y \cr 2b_i y & \de_{ij}} \ , \quad
\R'_{\mu \alpha} = \pmatrix { 1 & - 2b_j y' \cr 2b_i y' & \de_{ij}} \ .
\eqno (2.12) $$
Hence
$$
A_{i111}({\bxi}, y,y')= {2\over y^2-y'^2}\, \xi_i \, ( y\alpha -
y \beta - 2y' \gamma) + \rO ( \xi^2)
\eqno (2.13) $$
and the condition $\pr_i \langle T_{i1} T_{11} \rangle +
\pr_1 \langle T_{11} T_{11} \rangle = 0$ gives, using (2.10a),
$$
(y^2 - y'^2) \pr_y \beta = 2y' (d\beta - \gamma) \ .
\eqno (2.14) $$
Similarly
$$ \eqalign {
A_{i1k\ell}({\bxi}, y,y')={}& {2\over y^2-y'^2}\, \bigl ( \de_{k\ell}
\xi_i  ( y\beta - y \de ) - (\de_{ik} \xi_\ell + \de_{i\ell} \xi_k )
(y \epsilon - y' \gamma )\bigl ){}  + \rO ( \xi^2) \ , \cr
A_{ijk1}({\bxi}, y,y')={}& {2\over y^2-y'^2}\, \bigl ( \de_{ij}
\xi_k  ( y'\beta - y'\de ) + (\de_{jk} \xi_i + \de_{ik} \xi_j )
(y \gamma - y' \epsilon )\bigl ){}  + \rO ( \xi^2) \ , \cr}
\eqno (2.15) $$
and the associated conservation equations, using (2.10a,b), lead to
(2.14) again and also
$$
(y^2 - y'^2) \pr_y \gamma = 2y' (d\gamma - \beta + \de + d\epsilon ) \ .
\eqno (2.16) $$
(2.14) and (2.16) exhaust the restrictions following from restricted
conformal invariance and the conservation equation for $T_{\mu \nu}$
in the perpendicular geometry assumed for (2.8). In terms of the variable
$v$ in (2.9) (2.14) and (2.16) simplify to
$$
v{d\over dv} \beta = d\beta -2\gamma \ , \quad
v{d\over dv} \gamma = d\gamma - \beta + \de + d\epsilon \ .
\eqno (2.17) $$
Even after using (2.10a,b) to eliminate $\alpha$ and, for instance,
$\epsilon$ there remain two coupled linear differential equations
amongst three unknowns so in general an arbitrary function of $v$ appears
in the solution. This is exemplified by particular cases subsequently.
At large distances from the boundary the solutions should tend smoothly to
$$
\alpha(0) = (d-1)C \ , \quad \beta(0) = \de (0) = -C \ , \quad
-\gamma (0) = \epsilon (0) = \half d C \ .
\eqno (2.18) $$
For $d=2$ $C$ is proportional to the Virasoro central charge while
for $d=4$ $C$ is related to the coefficient of the Weyl tensor squared
in the trace of the energy momentum tensor on a curved space background [6].

However for $d=2$ the index $i$ is restricted to just $i=2$ and in (2.8)
$A_{\mu \nu \si \rho}$ depends only on the combination $\de + 2 \epsilon$.
This is reflected in eqs. (2.10b) and (2.17) so that they can be solved
to give
$$ \alpha = - \beta = \de + 2 \epsilon = C (1+ v^4) \ , \quad
\gamma = - C(1-v^4) \ .
\eqno (2.19) $$
The form of $\gamma$ corresponds to $T_{12}=0$ on the boundary $x_1=0$.

As a further application we may consider the correlation function of
the energy momentum tensor with a scalar field $\O$ of dimension $\eta$.
In this case as well as (2.6) for $\tx \to x$
$$ \O (\tx) \to \Omega(\tx)^\eta \O (\tx)
$$
and instead of (2.7)
$$
\langle T_{\mu \nu}(x) \O (x') \rangle = {1\over (x-x')^{d+\eta}} \,
S_{\mu \nu} ({\bf s},x_1, x'_1) \ .
\eqno (2.20) $$
In the perpendicular geometry $x_\mu=(y,{\bf 0}), \, x'_\mu=(y',{\bf 0}), \
y>y'$, then imposing tracelessness gives
$$
S_{11} = w(v) \ , \quad S_{ij} = - {1\over d-1} \, w(v) \, \de_{ij} \ .
\eqno (2.21) $$
Following a similar discussion to before
$$ S_{i1}({\bxi},y,y') = 2\xi_i \, {y\over y^2-y'^2} \, {d\over d-1}
\, w(v) + \rO (\xi^2) $$
and the conservation equation $\pr_i T_{i1} + \pr_1 T_{11}=0$ leads to
$$
{d\over dv} w(v) + {d-\eta\over 1-v}\, w - {\eta\over v} \, w =0 \ .
\eqno (2.22) $$
This has the solution
$$
w(v) = S \, v^\eta (1-v)^{d-\eta} \ , \quad y>y' \ .
\eqno (2.23) $$
Conversely if $y'>y$ then, with now $v=(y'-y)/(y+y')$, the solution
becomes
$$
w(v) = S \, v^\eta (1+v)^{d-\eta}  \ .
\eqno (2.24) $$

The coefficient $S$ appearing in (2.23,24) may be related to
$\langle \O(x)\rangle$ by using the operator product expansion of
$T_{\mu \nu}(x)$ and $\O(x')$ for $x\to x'$ which with our normalisations
has the form
$$ \eqalign {
& T_{\mu \nu}(y,{\bf 0}) \O(y',{\bf 0}) \sim {1\over |y-y'|^d} \,
A_{\mu \nu} \, \O(y',{\bf 0}) \ , \cr
& A_{11} = - A_{ii} \ , \quad A_{ij} = - {\eta\over S_d} \, \de_{ij} \ ,
\quad S_d = {2\pi^{\hh d}\over \Gamma(\hh d)} \ . \cr}
\eqno (2.25) $$
Hence from (2.20,21) and (2.23) or (2.24) with $v\to 0$
$$
\langle \O (y,{\bf 0})\rangle  = - {S_d\over \eta}\, {S\over (2y)^\eta} \ .
\eqno (2.26) $$

Cardy [3] derived further relations by considering analogous short distance
expansions in the neighbourhood of the boundary where $\O(x)$ is expanded
in terms of low dimension boundary operators. Supposing that the expansion
is restricted to the unit operator and the energy momentum operator
$T_{\mu \nu}$ itself on the boundary (where $T_{i1} = 0$) he assumed
$$
\O (y,{\bf 0}) \sim \langle \O (y,{\bf 0}) \rangle \bigl ( 1 + b \,
y^d T_{11}(0,{\bf 0}) + \dots \bigl) \ ,
\eqno (2.27) $$
then from the solution (2.25) for $y'\to 0$ and also (2.7,8)
$$
\langle T_{11} (y,{\bf 0})  T_{11} (0,{\bf 0}) \rangle = \alpha (1) \,
{1\over y^{2d}} \ , \quad \alpha (1) = - 2^d \, {\eta\over S_d} \,
{1\over b} \ .
\eqno (2.28) $$
This shows that the coefficient $b$ does not depend on the particular
operator $\O$.
In two dimensions from the solutions (2.19) $\alpha (1) = 2C$ and this leads
to so called hyperscaling relations [3] for $b$ which is then independent
of particular boundary conditions and which appears in universal terms
in the expression for the Casimir energy for parallel plate geometries.
For $d>2$ there appears to be no general relation between $\alpha(0)$
and $\alpha(1)$.

Alternatively if in (2.20) with the solution (2.24) we take $y\to 0$
then in this case the limit is non singular and from (2.26) we obtain
$$
\langle T_{11} (0,{\bf 0}) \O (y',{\bf 0}) \rangle = - {\eta\over S_d} \,
{2^d\over y'^d} \, \langle \O (y',{\bf 0}) \rangle \ .
\eqno (2.29) $$

More general configurations than the perpendicular geometry discussed
above may be obtained by conformal transformation. If $x'_\mu = (y',
{\bf 0})$ is fixed then we may take, generalising the infinitesimal
transformation (2.11),
$$
(y,{\bf 0}) \to (x_1 , \bx) \ , \quad x_1 = {1+ b^2 y'^2 \over1+ b^2 y^2}
\, y \ , \quad \bx =  {y^2 - y'^2 \over 1+ b^2 y^2} \, {\bf b}
\eqno (2.30) $$
and
$$
\R_{\mu \alpha} = \pmatrix { {1 - b^2 y^2 \over1+ b^2 y^2} &
- { 2b_j y \over 1+ b^2 y^2} \cr {2b_i y \over 1+ b^2 y^2} &
\de_{ij}  - {2 b_i b_j y^2 \over 1+ b^2 y^2}} \ ,
\eqno (2.31) $$
and correspondingly for $\R'_{\mu \alpha}$ with $y\to y'$. As the magnitude
of $\bf b$ varies $x_\mu$ moves on a semi-circle from $(y,{\bf 0})$ to
$(y'^2/y,{\bf 0})$ for $b^2\to \infty$, as shown in fig. 1, and $v\to -v$.
It is easy to
verify from (2.30) that $(y-y')^2\to (x-x')^2$ and $v^2= (y-y')^2/(y+y')^2
\to v^2$ as given by (2.4). Hence $\langle T_{\mu \nu}(x) T_{\si \rho}(x')
\rangle$ may be found for arbitrary $x,x'$ in terms of $\alpha,\beta,
\gamma,\de,\epsilon$. Assuming (2.6) and (2.7,8) with (2.30,31), taking $x'_1
= y'$, gives
$$ \eqalign {
\langle T_{\mu \nu}(x) T_{\si \rho}(x') \rangle = {1\over s^{2d}} \,
\Bigl \{& \de(v) \, \de_{\mu \nu} \de_{\si \rho} + \epsilon (v) \bigl (
I_{\mu \si} (s) I_{\nu \rho}(s) + I_{\mu \rho} (s) I_{\nu \si}(s) \bigl ) \cr
& + \bigl ( \beta(v) - \delta (v) \bigl ) \bigl ( X_\mu X_\nu \,
\de_{\si \rho} + X'{}_{\! \si} X'{}_{\! \rho} \, \de_{\mu \nu} \bigl ) \cr
& - \bigl ( \gamma(v) + \epsilon (v) \bigl ) \bigl ( X_\mu X'{}_{\! \si} \,
I_{\nu \rho}(s) + X_{\nu} X'{}_{\! \rho} \, I_{\mu \si}(s) \cr
& ~~~~~~~~~~~~~~~~~~~ + X_{\mu} X'{}_{\! \rho} \, I_{\nu \si} (s) +
X_{\nu} X'{}_{\! \si} \, I_{\mu \rho} (s) \bigl ) \cr
& + \bigl ( \alpha(v) - 2\beta (v) + 4\gamma(v) + \de (v) + 2\epsilon (v)
\bigl ) X_\mu X_\nu X' {}_{\! \si} X' {}_{\! \rho} \Bigl \} \ , \cr}
\eqno (2.32) $$
where $v$ is given by (2.4), $s=x-x'$ and
$$ \eqalign {
I_{\mu \si}(s) = {}& \de_{\mu \si} - 2\, {s_\mu s_\si \over s^2} \ , \cr
X_\mu = {}& \R_{\mu \alpha} n_\alpha = N \bigl ( x_1^{\, 2} -
x_1^{\prime \, 2} - {\bf s}^2 , \, 2x_1^{\vphantom +} \, {\bf s} \bigl
) \ , \cr
X'{}_{\! \si} = {}& - \R'{}_{\! \si \alpha} n_\alpha = I_{\si \mu} (s) X_\mu
= N \bigl ( x_1^{\prime \, 2} - x_1^{\, 2} - {\bf
s}^2 , \, - 2x_1^{\prime \vphantom +} \, {\bf s} \bigl ) \ , \cr
N^{-2} = {}& s^2 ( s^2 + 4x_1 x'{}_{\! 1} ) = (s^2)^2/v^2 \ , \cr}
\eqno (2.33) $$
for $n_\alpha = (1,{\bf 0} )$ defining the normal to the boundary.
It is easy to check that (2.32) is consistent
with $T_{\mu \mu} =0$, using (2.10a,b), since $X,X'$ are unit vectors.
At large distances from the boundary, when $v\to
0$, all terms containing $X$ or $X'$ vanish by virtue of (2.18) and the usual
result [6] for no boundary is obtained.

As a special case we may obtain
$$
\langle T_{11} (0,\bx) T_{11} (0,\bx') \rangle =  \alpha (1) \,
{1\over (\bx - \bx')^{2d}} \ ,
\eqno (2.34) $$
which may also be found more directly from (2.28). In a similar fashion from
(2.29)
$$
\langle T_{11} (0,{\bf 0}) \O (y,\bx) \rangle =  - {\eta\over
S_d} \, \Bigl ( {2y \over \bx^2 + y^2} \Bigl )^d \langle \O (y,{\bx})
\rangle \ .
\eqno (2.35) $$
\bigskip
\leftline{\bigbf 3 Calculations in Specific Models}
\medskip

The simplest conformal field theory for arbitrary $d$ is perhaps that
corresponding to a free scalar field $\phi$ for which the energy momentum
tensor is
$$
T_{\mu \nu} = \pr_\mu \phi \pr_\nu \phi - \quar \, {1\over d-1} \bigl (
(d-2) \pr_\mu \pr_\nu + \de_{\mu \nu} \pr^2 \bigl ) \phi^2 \ .
\eqno (3.1) $$
This is conserved and traceless on the equations of motion $\pr^2\phi =0$.
In order to maintain conformal invariance with a plane boundary at $x_1=0$
it is sufficient to impose Neumann $\pr_1 \phi (0,{\bf 0})=0$ or Dirichlet
$\phi (0,{\bf 0})=0$ boundary conditions. In the perpendicular
geometry described in the previous section the basic two point function
is
$$
\langle  \phi (y,{\bf 0}) \phi (y',{\bf 0}) \rangle = {1\over S_d (d-2)}
\Bigl ( {1\over |y-y'|^{d-2}} \pm {1\over (y+y')^{d-2}} \Bigl)
\eqno (3.2) $$
and we may easily find
$$ \eqalign {
\langle \pr_i \phi (y,{\bf 0}) \pr_j \phi (y',{\bf 0}) \rangle = {}&
- \langle  \pr_i \pr_j \phi (y,{\bf 0}) \phi (y',{\bf 0}) \rangle =
{1\over S_d}\, \de_{ij}
\Bigl ( {1\over |y-y'|^{d}} \pm {1\over (y+y')^{d}} \Bigl)\ , \cr
\langle  \pr_i \pr_j \phi (y,{\bf 0}) \pr_k \pr_\ell\phi (y',{\bf 0})
\rangle = {}& {d\over S_d}\, \bigl ( \de_{ij}\de_{k\ell} + \de_{ik}
\de_{j\ell} + \de_{i\ell} \de_{jk} \bigl )
\Bigl ( {1\over |y-y'|^{d+2}} \pm {1\over (y+y')^{d+2}} \Bigl)\ . \cr}
\eqno (3.3) $$
With these results it is a matter of straightforward calculation to
find
$$\eqalign {
S_d^2 \alpha(v) = {}& 1 + v^{2d} \pm \quar (d-2)d\, {d+1\over d-1}\,
v^{d-2}(1-v^2)^2 \ , \cr
S_d^2 \gamma(v) = {}& - {d\over 2(d-1)} \Bigl (
1 - v^{2d} \pm \half (d-2) {d+1\over d-1}\,
v^{d-2}(1-v^4)\Bigl )  \ , \cr
S_d^2 \epsilon(v) = {}&  {d\over 2(d-1)} \bigl (
1 + v^{2d}\bigl ) {} \pm \quar\, {d\over (d-1)^2}\, \bigl (
(d-2) (v^{d-2}+v^{d+2}) +2d\, v^d \bigl ) \ , \cr}
\eqno (3.4) $$
with $\beta,\de$ determined from (2.10a,b). It is easy to verify that
these results satisfy the differential equations (2.17). For $v\to 0$
(3.4) gives $C=(d-1)^{-1}S_d^{-2}$ in (2.18).

For the scalar operator $\phi^2$, of dimension $d-2$ in this free theory,
then from (3.2) after subtraction of the usual short distance divergence
$$
\langle \phi^2 (y,{\bf 0}) \rangle = \pm {1\over (d-2) S_d} \, {1\over
(2y)^{d-2}} \ ,
\eqno (3.5) $$
and it is easy to verify that this is compatible with
$\langle T_{\mu \nu}(y,{\bf 0}) \phi^2 (y',{\bf 0}) \rangle$ as determined
from (2.20,21) and (2.23) or (2.24).

An equally simple conformal field theory is that provided by free fermion
fields for which the energy momentum tensor is
$$ T_{\mu \nu} = \half {\bar \psi} ( \gamma_\mu {\olr \pr}_\nu + \gamma_\nu
{\olr \pr}_\mu ) \psi \ .
\eqno (3.6) $$
The basic two point function is taken as
$$ \langle \psi(x) {\bar \psi(x')} \rangle = {1\over S_d} \Bigl (
{\gamma {\cdot (x-x')}\over (x-x')^d} + U \, {\gamma {\cdot ({\bar x}-x')}
\over ({\bar x}-x')^d} \Bigl ) \ , \quad {\bar x}_\mu = (-x_1, \bx) \,
\eqno (3.7) $$
where $U$ is a matrix satisfying
$$ U \gamma_1 = - \gamma_1 {\bar U} \ , \quad U \gamma_i = \gamma_i {\bar U}
\ , \quad U^2 = {\bar U}^2 = 1 \ .
\eqno (3.8) $$
This corresponds to boundary conditions
$$ (1-U) \psi \bigl |_{\pr \M} = 0 \ , \quad
{\bar \psi} (1- {\bar U} )\bigl | _{\pr \M} = 0 \ .
\eqno (3.9) $$
It is straightforward to show that in this case
$$ \epsilon(v) = - \half d \de (v) \ , \quad \beta(v) = \de(v) = - \half \,
{2^{\hh d}\over S_d^2 } \bigl ( 1+v^{2d} \bigl ) \ , \quad
\gamma(v) = - \quar d \,{2^{\hh d} \over S_d^2} \bigl ( 1-v^{2d} \bigl ) \ ,
\eqno (3.10) $$
which is similar to the scalar field case after averaging over Neumann and
Dirichlet boundary conditions.

For a less trivial calculation of the universal functions defining the two
point function of the energy momentum tensor we consider the well known
critical point in scalar field theories with a $\phi^4$ interaction
where $g_* = \rO (\vep)$. At this non Gaussian fixed point the theory is
conformally invariant and moreover critical exponents and other universal
quantities may be calculated as
an expansion in $\vep = 4-d$ using standard perturbative loop expansions.
Thus for $\phi $ an $n$-component field with an $O(n)$ invariant interaction
$$ V(\phi) = {\ts{1\over 24}} g \, (\phi^2)^2
\eqno (3.11) $$
we consider the $\rO (g)$ contributions to $\epsilon$ and $\gamma$.
These involve only off-diagonal components of $T_{\mu \nu}$ so that the
expression (3.1) is still sufficient. The relevant Feynman diagrams
are calculated in appendix C for either Neumann or Dirichlet
boundary conditions. The leading terms in the $\vep$ expansion are
given by (3.4) while the order $\rO (\vep)$ corrections from (C.7a,b) and
(C.9a,b) are
$$ \eqalignno {
\epsilon^{(1)}(v) = K \, \vep \, \Biggl\{&
\mp v^2(1+4v^2+v^4) \ln {(1-v^2)^2\over v^2}  \mp {\textstyle {9\over 2}}
v^2(1+v^4) \mp 2v^4  \cr
& + \bigl ( - 3(1+v^8) + {\textstyle {1\over 5}}v^2(1-v^2)^2 \bigl )
\ln {(1-v^2)^2\over v^2} \cr
& - 3(1-v^8) \ln v^2 + {\textstyle {1\over 5}}v^2(1-v^4) \Bigl (
1 - {2v^2\over (1-v^2)^2} + {6v^4\over (1-v^2)^4} \Bigl ) \ln v^2 \cr
& -5v^2(1+v^4) - {\textstyle {8\over 5}}v^4 + {\textstyle {12\over 5}}\,
{v^6 \over (1-v^2)^2}\Biggl\}  \ , & (3.12a) \cr
\gamma^{(1)}(v) = K \, \vep \, \Biggl\{&
\pm 5v^2(1- v^4) \Bigl ( \ln {(1-v^2)^2\over v^2} - {1\over 2} \Bigl )\cr
& + \bigl ( 3(1-v^8) - v^2(1-v^4) \bigl ) \ln {(1-v^2)^2\over v^2} \cr
& + \bigl ( 3(1+v^8) - v^2(1+v^4) \bigl )\ln v^2 \cr
& + {2v^6\over (1-v^2)^2} \, \ln v^2
+ {v^2(1+v^6)\over 1-v^2} \Biggl\}
 \ , & (3.12b) \cr}
$$
where
$$ K = {1\over S_4^2}\, {n\over 9}\,  {n+2 \over n+8} \ .
\eqno (3.13) $$
It is straightforward to verify that these results satisfy (2.17), together
with (2.10a,b), which in this context become
$$
\Bigl ( v{d\over dv}- 4\Bigl ) \gamma^{(1)}  = - {\textstyle {4\over 3}}\,
\beta^{(1)} + {\textstyle {10\over 3}}\, \epsilon^{(1)} \ , \quad
\Bigl ( v{d\over dv}- 4 \Bigl ) \beta^{(1)} = -2\gamma^{(1)} \ .
\eqno (3.14) $$
{}From these equations
$$ \eqalign {
\beta^{(1)}(v) = K \, \vep \, \Biggl\{&
\pm 5 v^2(1-v^2)^2 \ln {(1-v^2)^2\over v^2}  \mp {\textstyle {15\over 2}}
v^2(1+v^4) \pm 10 v^4  \cr
& + {\textstyle {1\over 2}}(1-v^2)^2(3+4v^2 + 3v^4)
\ln {(1-v^2)^2\over v^2} \cr
& +{\textstyle {3\over 2}}(1-v^8) \ln v^2
- {v^2(1+v^6)\over 1-v^2} \, \ln v^2 \cr
& -2v^2(1+v^4) + 2v^4 \Biggl\}  \ . \cr}
\eqno (3.15) $$
All quantities $\epsilon^{(1)},\, \gamma^{(1)}, \, \beta^{(1)}$ vanish as
$v\to 0$ since there is no change in $C$, given by (2.18), at this order.
As required by the boundary condition $T_{i1} = 0$ for $x_1=0$ $\gamma^{(1)}
(1) = 0$. However $\epsilon^{(1)}(v)$ is singular as $v\to 1$ although
$\beta^{(1)}(v)$ is well behaved in this limit. In the next
section it is shown that $T_{11} = - T_{ii}$ is a well defined operator on
the boundary $x_1=0$ but this does not restrict the traceless part of
$T_{ij}$ to be finite in general on the boundary.
Of course for $d=2$ all components of $T_{\mu \nu}$ are non singular near the
boundary. From the above expressions for $\beta$ for $v=1$ it is easy to see
that the leading terms in the $\vep$ expansion give
$$ \alpha(1) = {n\over S_d^2} \Bigl ( 2 \pm {5\over 3} \, {n+2\over n+8}
\, \vep \Bigl ) \ ,
\eqno (3.16) $$
which coincides exactly with the results of Eisenriegler {\it et al.} [4] who
calculated the two point function of $T_{11}$ at $x_1=0$ as in (2.34).

\bigskip
\leftline{\bigbf 4 Energy Momentum Tensor on Curved Manifolds with Boundary}
\medskip

In a quantum field theory defined on a manifold $\M$ with an arbitrary
smooth metric $g_{\mu \nu}$ the energy momentum tensor may be defined
as a finite local composite operator by considering variations with respect
to $g_{\mu \nu}$. For $x^\mu\in \M$ then if $\M$, of dimension $d$, has
a boundary $\pr \M$, of dimension $d-1$, this may be determined by
$x_\b^\mu (\hbx)$ for $\hbx$ arbitrary coordinates on $\pr \M$. On $\pr \M$
the induced metric is defined by
$$
\hga_{ij}(\hbx) = g_{\mu \nu}(x_\b)  e^\mu {}_{\! i}(\hbx) e^\nu {}_{\! j}
(\hbx) \ , \quad
 e^\mu {}_{\! i} (\hbx ) = {\pr x_\b^\mu \over \pr \hx^i} ~ ,
\eqno (4.1) $$
and we may also define related geometric quantities such as the unit
inward normal $n^\mu(\hbx)$, satisfying $ n_\mu
e^\mu {}_{\! i} = 0, \, n_\mu n^\mu = 1 $, and the extrinsic curvature
$K_{ij} (\hbx)$ given by (A.1) which has dimension 1.

For any quantum field theory the vacuum energy functional $W(g,x_\b)$,
which here depends on the metric $g_{\mu \nu}$ and the embedding of $\pr \M$
in $\M$ given by $x_\b$, may be defined by a functional integral of the
generic form
$$ e^W = \int \! d[\phi] \, e^{-S_0(\phi)}
\eqno (4.2) $$
for fields $\phi$ on $\M$ satisfying suitable boundary conditions on
$\pr \M$ and $S_0$ the bare action including all necessary
counterterms in an appropriate regularisation scheme so that
$W$ is a finite functional for arbitrary smooth metrics
$g_{\mu \nu}$ and boundaries $x_\b^\mu$. Recently [7] we have shown how this
may be achieved to two loops using dimensional regularisation in four
dimensions for renormalisable scalar field theories with quantum fields
obeying either Dirichlet or generalised Neumann boundary conditions.

In such a framework we may write under variations in the metric
$$ \eqalign {
- \de_g W = {}& \int_\M \! dv \, \half \de g^{\mu \nu} \langle T_{\mu \nu}
\rangle \cr
& + \int_{\pr \M} \! \! \! dS \, \bigl ( \half \de \hga^{ij} \langle
B_{ij} \rangle + \de n^\mu \langle \lambda_\mu \rangle + \half \de
K_{ij} \langle C^{ij} \rangle + \dots \bigl ) \ , \cr}
\eqno (4.3) $$
where $dv=d^dx\sqrt g , \, dS=d^{d-1}\hx \sqrt \hga$. This relation
defines insertions of local operators $B_{ij}(\hbx), \, \lambda_n (\hbx),
\, \lambda_i(\hbx)$ and $C^{ij}(\hbx)$ on $\pr \M$. The neglected surface
terms in (4.3) involve variations of higher dimension geometric tensors on
$\pr \M$, such as components of the curvature tensor like $R_{ninj} =
n^\mu e^\nu {}_{\! i}\,  n^\si e^\rho {}_{\! j}\, R_{\mu \nu \si \rho}$, but
these are absent in simple theories.

To derive conservation equations we assume invariance under diffeomorphisms,
or reparameterisations of the coordinates, as given by $\de_v x^\mu = -
v^\mu (x)$. On the boundary $\pr \M$ we also require $\de_v \hx^i = -
v^i (\hbx)$ where
$ v^\mu (x_\b) = e^\mu {}_{\! i}(\hbx) v^i (\hbx) + n^\mu (\hbx) v_n (\hbx)$.
On $\M$ the external metric is required to transform as
$$ \de_v g_{\mu \nu} = \nab_\mu v_\nu + \nab_\nu v_\mu \ .
\eqno (4.4) $$
The corresponding variations on $\pr \M$ in $\hga_{ij} , \, n^\mu $
and $K_{ij}$ induced by $\de_v g_{\mu \nu}$ in (4.4) are obtained in
the appendix to be
$$ \eqalign {
\de_v \hga_{ij} = {}& \hnab_i v_j + \hnab_j v_i - 2 K_{ij} \, v_n \ , \cr
\de_v n^\mu = {}& - \nab_n v^\mu - e^{\mu i} (\pr_i v_n + K_{ij} v^j ) \ , \cr
\de_v K_{ij} = {}& \L_v K_{ij} + \bigl ( R_{ninj} - K_{ik} K_j{}^k \bigl)
v_n + \hnab_i \pr_j v_n \ , \cr
& \L_v K_{ij} = v^k \hnab_k K_{ij} + \hnab_i v^k K_{kj} + \hnab_j v^k
K_{ik} \ , \cr }
\eqno (4.5)$$
for $\hnab_i$ the covariant derivative acting on tensor fields over $\pr \M$
with the Christoffel connection prescribed by the metric $\hga_{ij}$. For
a general diffeomorphism it is necessary to assume also a shift in the
boundary surface represented by $\de_v x_\b^\mu (\hbx) = - n^\mu (\hbx)
v_n(\hbx)$.

By integration by parts invariance under diffeomorphisms on $\M$ gives the
usual conservation equation
$$ \nab^\mu \langle T_{\mu \nu} \rangle = 0 \ ,
\eqno (4.6)$$
(for simplicity we neglect contributions from any sources for other local
operators on which $W$ may also depend). On the boundary invariance gives
from the results (4.5)
$$ \eqalign {
\langle \lambda_\mu \rangle = {}& 0 \ , \cr
\langle T_{ni} \rangle = {}& - \hnab^j \langle B_{ij} \rangle - \half
\hnab_i K_{jk} \, \langle C^{jk} \rangle + \hnab_j \bigl ( K_{ik} \langle
C^{jk} \rangle \bigl ) \ . \cr}
\eqno (4.7) $$
In addition, after setting $\langle \lambda_\mu \rangle = 0$, we may also
obtain
$$
-n^\mu {\de \over \de x_\b^\mu}W = \langle T_{nn} \rangle + K^{ij}
\langle B_{ij} \rangle + \half \bigl (  R_{ninj} - K_{ik} K_j{}^k \bigl)
\langle C^{ij} \rangle + \half \hnab_i \hnab_j \langle C^{ij} \rangle \ .
\eqno (4.8) $$
(4.7) and (4.8) ensure that $T_{nn} = n^\mu n^\nu T_{\mu \nu}$ and
$T_{ni} = n^\mu e^\nu {}_{\! i}\,  T_{\mu \nu}$ are well defined local
operators on $\pr \M$.

If we also assume Weyl invariance under local rescalings of the metric
$$
\de_\si g^{\mu \nu} = 2\si \, g^{\mu \nu} \,
\eqno (4.9) $$
then on $\M$ this requires as usual
$$ g^{\mu \nu} \langle T_{\mu \nu} \rangle = 0 \ .
\eqno (4.10) $$
{}From (4.9) the induced variations on the boundary are
$$
\de_\si \hga^{ij} = 2\si \, \hga^{ij} \ , \quad \de_\si n^\mu = \si \, n^\mu
\ , \quad \de_\si K_{ij} = \si \, K_{ij} + \hga_{ij} \pr_n \si \ .
\eqno (4.11) $$
Weyl invariance then also requires
$$ \eqalign {
& \hga^{ij} \langle B_{ij} \rangle + \half K_{ij} \langle C^{ij} \rangle
+ \langle \lambda_n \rangle = 0 \ , \cr
& \hga_{ij} \langle C^{ij} \rangle = 0 \ . \cr}
\eqno (4.12) $$
For $d=2$ and a plane boundary $x_1=0$ (4.7) is equivalent to the operator
relation on the boundary $T_{1x} = - \pr_x B$, for $x_2=x$, while from
(4.12) for conformal invariance $B=0$\footnote{*}{Similar equations have
been obtained by Cardy [8].}.

As an illustration of these results we a consider a simple scalar field
theory specified by bulk and surface contributions to the action of the
form
$$ \eqalign {
S(\phi) = {}& \int_\M \! dv \, \L \ , \quad \L = \half \pr_\mu \phi
\pr^\mu \phi + \half \tau R \,  \phi^2 + V(\phi) \ , \cr
{\hat S} (\phi) = {}& \int_{\pr \M} \! \! \! dS \, \bigl ( Q(\phi) +
\half \rho K \, \phi^2 \bigl ) \ , \cr}
\eqno (4.13) $$
where $R$ is the scalar curvature on $\M$ and $K=\hga^{ij} K_{ij}$. In the
corresponding quantum field theory to lowest order in the loop expansion
we may write
$$ W(g,x_\b)^{(0)} = - S(\phi) - {\hat S}(\phi) \ ,
\eqno (4.14) $$
where $\phi$ is a solution of the classical field equations and boundary
conditions such that $S+{\hat S}$ is stationary,
$$
\nab^2\phi - \tau R \, \phi - V'(\phi) = 0 \ , \quad
\bigl ( \pr_n \phi - \rho K \,\phi - Q'(\phi) \bigl ) \bigl | _{\pr \M} = 0 \ .
\eqno (4.15) $$

Under a variation in the metric, since $\de_g R = \de g^{\mu \nu} R_{\mu \nu}
- (\nab_\mu \nab_\nu - g_{\mu \nu} \nab^2 ) \de g^{\mu \nu}$,
$$ \eqalign {
\de_g S(\phi) = {}& \int_\M \! dv \, \half \de g^{\mu \nu} T_{\mu \nu} \cr
+ &\int_{\pr \M} \! \! \! dS \, \half \tau \bigl ( (- h_{\mu \nu}\nab_n
\de g^{\mu \nu} + n_\mu e_\nu {}^{\! i}\nab_i \de g^{\mu \nu})\phi^2
+ h_{\mu \nu} \de g^{\mu \nu}\pr_n \phi^2 -n_\mu e_\nu {}^{\! i}
\de g^{\mu \nu}\pr_i \phi^2 \bigl ) \, , \cr}
\eqno (4.16) $$
for $h_{\mu \nu} = g_{\mu \nu} - n_\mu n_\nu$ and
$$
T_{\mu \nu} = \pr_\mu \phi \pr_\nu \phi + \tau R_{\mu \nu} \, \phi^2
- \tau (\nab_\mu \nab_\nu - g_{\mu \nu} \nab^2) \phi^2 - g_{\mu \nu}\, \L \ .
\eqno (4.17) $$
It is easy to check that $\nab^\mu T_{\mu \nu} = 0$ using the equation of
motion (4.15). For $V=0$ and $\tau = \quar (d-2)/(d-1)$ it is also
straightforward to verify that $g^{\mu \nu} T_{\mu \nu} = 0$ on flat space and
that $T_{\mu \nu}$ is then equivalent to that assumed in (3.1) for the free
scalar conformal field theory.
Using (4.16) and the results in the appendix it is possible to write
$$
\de_g (S+{\hat S}) = \int_\M \! dv \, \half \de g^{\mu \nu} T_{\mu \nu}
+ \int_{\pr \M} \! \! \! dS \, \bigl ( \half \de \hga^{ij}
B_{ij} + \de n^\mu  \lambda_\mu  + \half \de
K_{ij}  C^{ij} \bigl ) \ ,
\eqno (4.18) $$
where, using the boundary conditions on $\phi$ in (4.15),
$$ \eqalign {
B_{ij} = {}& \hga_{ij}\bigl ( - Q(\phi) +2\tau Q'(\phi)\phi
+ 2\tau\rho K \phi^2 - \half \rho K
\phi^2 \bigl ){} + (\rho - \tau ) K_{ij} \, \phi^2 \ , \cr
C^{ij} = {}& (\rho - 2\tau )\hga^{ij} \, \phi^2 \ , \quad \quad \quad
\lambda_\mu = 0 \ . \cr}
\eqno (4.19) $$

The general results derived above may now be verified directly.
On the boundary from (4.17)
$$T_{ni} = \pr_n \phi \pr_i \phi + \tau R_{ni} \, \phi^2
- \tau \, \pr_i \pr_n \phi^2 - \tau \, K_i {}^{\! j}\pr_j \phi^2 \ ,
\eqno (4.20) $$
where $\pr_n \phi$ may be eliminated from (4.15). Using the Gauss-Codazzi
equation $R_{ni} = \pr_i K - \hnab_j K_i {}^{\! j}$ it is then easy to
show that
$$
T_{ni} = - \hnab^j B_{ij} - \half \hnab_i K_{jk} \, C^{jk} + \hnab_j
(K_{ik} C^{jk} ) \ ,
\eqno (4.21) $$
which is in accord with (4.7) as expected. On the boundary also
$$T_{nn} = \pr_n \phi \pr_n \phi - \half \pr^i \phi \pr_i \phi - \half \tau
R \phi^2 - V(\phi) + \tau R_{nn} \phi^2 + \tau \hnab^2 \phi^2
- \tau K \pr_n \phi^2  \ ,
\eqno (4.22) $$
and using (A.7,8,9) we may verify (4.8) in this case as well
$$
n^\mu {\de \over \de x_\b^\mu}(S+{\hat S}) = T_{nn} + K^{ij}
B_{ij} + \half \bigl (  R_{ninj} - K_{ik} K_j{}^k \bigl)
C^{ij} + \half \hnab_i \hnab_j  C^{ij} \ .
\eqno (4.23) $$

For Weyl invariance from the second of the conditions in (4.12) $\hga_{ij}
C^{ij}=0$ we obtain $\rho = 2\tau$ and then
$$
\hga^{ij} B_{ij} = (d-1) \bigl ( - Q(\phi) + 2\tau Q'(\phi)\phi \bigl ){}
+ \tau \bigl ( 4(d-1) \tau - (d-2) \bigl) K\phi^2 \ .
\eqno (4.24) $$
Hence $ \hga^{ij} B_{ij} = 0$ gives the same result for $\tau$ as $g^{\mu \nu}
T_{\mu \nu} = 0$. If $Q(\phi) = \half c \phi^2$ the remaining condition is
satisfied if either $c=0$, and $\pr_n \phi |_{\pr \M} = 0$, or if $c\to
\infty,\ c\phi = \rO (1)$, so that $\phi|_{\pr \M} = 0$, showing how both
Neumann and Dirichlet boundary conditions on scalar fields are separately
compatible with conformal invariance.

\bigskip
\leftline{\bigbf 5 Conclusion}
\medskip

It is clear from the results of this paper that two-point correlation
functions of
the energy momentum tensor in conformal field theories with appropriate
boundary conditions preserving conformal invariance may have a dependence
on the conformal invariant variable $v$, defined here by (2.4), which depends
on the particular conformally invariant theory and its boundary conditions
whenever $d>2$. An interesting question, beyond the scope of our considerations
here, is whether such functions are measurable in realistic statistical
physics models at a critical point when we may take $d=3$. For models in
the same universality class as the Ising model then our $\vep$ expansion
results would perhaps be relevant setting $\vep=1$. For such applications
it would be desirable to find functions which agreed with our results to
$\rO(\vep)$ but remained solutions of the equations (2.10a,b) and (2.17)
for general $d$ since they would then extrapolate to the unique functional
form for $d=2$ given by (2.19).
\bigskip

We are grateful to John Cardy for stimulating conversations and sending us
a copy of ref. 4.
\vfill\eject
\leftline{\bigbf Appendix A}
\medskip

For application in section 4 we here summarise the essential results in a
geometrical treatment of a boundary $\pr \M$, parameterised by
coordinates $\hx^i$, where $\pr \M$ is specified in terms of the
coordinates for $\M$ by $x_\b^\mu(\hbx)$. A natural tangent frame basis
on $\pr \M$ is given by $e^\mu {}_{\! i} (\hbx ), \, n^\mu(\hbx)$, with
$n^\mu$ the unit inward normal, and the extrinsic metric on $\pr \M$ is
defined as in (4.1). The symmetric tensor $K_{ij}$ forming the extrinsic
curvature and a connection ${\hat \Gamma}^k_{ij}$ are defined on $\pr \M$ by
$$\eqalign {
\nab_i  e^\mu {}_{\! j}  = {}& \pr_i e^\mu {}_{\! j}  + \Gamma^\mu_{\si \rho}
e^\si {}_{\! i} e^\rho {}_{\! j} -  e^\mu {}_{\! k}
{\hat \Gamma}^k_{ij} = n^\mu K_{ij} ~,\cr
\nab_i n^\mu = {}& \pr_i n^\mu   + \Gamma^\mu_{\si \rho}
e^\si {}_{\! i} n^\rho =
- K_{ij} e^{\mu j} ~ , ~~~~~~~~ \pr_i = e^\mu {}_{\! i} \pr_\mu \ , \cr}
\eqno(A.1) $$
where $\Gamma^{\mu}_{\si \rho}$ the usual Christoffel connection formed from
$g_{\mu \nu}$. Acting on tensors on $\pr \M$ a covariant derivative
$\hnab_i$ may be defined by the connection
${\hat {\Gamma}}^k_{ij} = {\hat {\Gamma}}^k_{ji}$ and from (A.1) and (4.1)
$\hnab_i \hga_{jk} = 0$ so that ${\hat {\Gamma}}^k_{ij}$
is just the Christoffel connection formed from $\hga_{jk}$. From (A.1) it is
straightforward to derive the Gauss-Codazzi equations relating the
Riemann curvature tensor $R_{\mu \nu \si \rho}$ for $x\in \pr \M$, with
zero and one component along the normal $n^\mu$, to the intrinsic
Riemann curvature ${\hat R}_{ijk\ell}$ of $\pr \M$ associated with the
covariant derivative $\hnab_i$ and also the extrinsic curvature $K_{ij}$.

To derive the implications for $\hga_{ij}, \, n^\mu$ and $K_{ij}$
of the variation of the basic metric (4.5)
induced by a diffeomorphism we first note that from (A.1)
$e^\mu {}_{\! i} e^\nu {}_{\! j}\nab_\mu v_\nu = \hnab_i v_j - K_{ij} v_n$
which leads directly from the definition (4.1) to $\de_v \hga_{ij}$ in
(4.6). The variation of the normal vector $n^\mu$ in (4.6) may be determined
from the equations $\de_v n_\mu e^\mu {}_{\! i} = 0$ and $\de_v n_\mu
n^\mu = - \half \de_v g^{\mu \nu} n_\mu n_\nu = n^\mu n^\nu \nab_\mu v_\nu$.
For the variation in the extrinsic curvature $K_{ij} = n_\mu \nab_i
e^\mu {}_{\! j}$ from (A.1) in order to obtain the result in (4.6) we use
$$ \eqalign {
\de_v K_{ij} = {}& n^\mu n^\nu \nab_\mu v_\nu \, K_{ij} + n_\mu \de_v
\Gamma^\mu_{\si \rho} \, e^\si {}_{\! i} e^\rho {}_{\! j} \ , \cr
\de_v \Gamma^\mu_{\si \rho} = {}&  g^{\mu \nu} \nab_{(\si} \nab_{\rho)} v_\nu
- v^\lambda R_{\lambda(\si \rho)}{}^\mu \ , \quad R_{k(ij)n} = \hnab_{(i}
K_{j)k} - \hnab_k K_{ij} \ , \cr
n_\mu e^\si {}_{\! i} e^\rho {}_{\! j}\nab_\si \nab_\rho v^\mu = {}&
- K_{ij} n^\mu n^\nu \nab_\mu v_\nu  + \hnab_i \pr_j v_n - K_{ik} K_j {}^k
v_n + \hnab_j v^k \, K_{ik} + \hnab_i (K_{jk} v^k) \ . \cr}
$$

To express the variations in the metric and its derivatives on $\pr \M$
in terms of variations in $\hga_{ij},\, n^\mu$ and $K_{ij}$, as required
for the simplification of (4.16), we use
$$ \de g^{\mu \nu} = \de n^\mu \, n^\nu + n^\mu\, \de n^\nu +
e^\mu {}_{\! i} e^\mu {}_{\! j} \de \hga^{ij} \ .
\eqno (A.2) $$
It is easy to see that, for $h_{\mu \nu} = g_{\mu \nu} - n_\mu n_\nu$,
$$
h_{\mu \nu} \de g^{\mu \nu} = \hga_{ij} \de \hga^{ij} \ , \quad
n_\mu e_\nu {}^{\! i} \de g^{\mu \nu} = e_\nu {}^{\! i}\de n^\nu \ ,
\eqno (A.3) $$
and also, using (A.1),
$$
n_\mu e_\nu {}^{\! i}\nab_i \de g^{\mu \nu} = \hnab_i
(e_\nu {}^{\! i}\de n^\nu) - 2K\, n_\mu \de n^\mu + K_{ij} \de \hga^{ij} \ .
\eqno (A.4) $$
To determine a corresponding expression for $h_{\mu \nu}\nab_n \de g^{\mu \nu}$
we first use $K_{ij} = - e^\mu {}_{\! i} e^\mu {}_{\! j} \nab_\mu n_\nu$ to
find
$$ \eqalign {
\de K_{ij} = {}& - e^\mu {}_{\! i} e^\mu {}_{\! j} \nab_\mu \, \de n_\nu +
e^\mu {}_{\! i} e^\mu {}_{\! j} n_\lambda \, \de \Gamma_{\mu \nu}^\lambda \cr
= {} & - e^\mu {}_{\! i} e^\mu {}_{\! j} \nab_\mu \, \de n_\nu -
e_{\mu (i} \hnab_{j)} \de n^\mu + K_{ij}\, n_\mu \de n^\mu -
\hga_{k(i} K_{j)\ell} \de \hga^{k\ell} + \half e_{\mu i} e_{\nu j} \nab_n
\de g^{\mu \nu} \ . \cr}
$$
{}From this it is straightforward to obtain
$$ h_{\mu \nu} \nab_n \de g^{\mu \nu} = 2 \de K - 2 K\, n_\mu \de n^\mu
+ 2 \hnab_i ( e_\mu {}^{\! i}\de n^\mu) \ .
\eqno (A.5) $$

Finally we consider the variations induced by a shift in the boundary surface
along a normal as given by $\de_t x_\b^\mu = - n^\mu \de t$. We let
$$ \eqalign {
\de'_t e^\mu {}_{\! i} ={}& \pr_i \de_t x_\b^\mu - \de t \, n^{\si}
\Gamma^\mu_{\si \nu}
e^\nu {}_{\! i} = \de t \, K_{ij} e^{\mu j} - n^\mu \pr_i \de t ~, \cr
\de'_t n_\mu = {}& \de_t n_\mu - n_\nu\Gamma^\nu_{\si \mu}n^\si \de t =
e_\mu {}^{\! i} \pr_i \de t ~ . \cr}
\eqno(A.6)
$$
With the induced metric given by (4.1) we have
$$
\de_t \hga_{ij} = 2 K_{ij} \, \de t ~, \eqno(A.7)
$$
and using $K_{ij} = n_\mu \nab_i e^\mu{}_{\! j}$ from (A.1)
$$\eqalign {
\de_t K_{ij} = {}& e_\nu {}^{\! k} \pr_k \de t \, \nab_i e^\nu {}_{\!
j} - n_\mu
[\nab_n,\nab_i] e^\mu {}_{\! j} + n_\mu \nab_i \de'_t e^\mu {}_{\! j} \cr
= {}& -\hnab_i \pr_j \de t - \bigl ( R_{njni} - K_{jk} K^k {}_{\! i}\bigl )
\de t ~. \cr}
\eqno(A.8)
$$
For integrals over a local scalar function $f$ on $\M$ or, restricted
to the boundary, on $\pr \M$
$$\eqalign {
 \de_t \int_\M \!  dv\, f ={}& \int_{\pr \M} \! \!  dS\, \de t \, f |_{\pr
\M} ~, \cr
\de_t \int_{\pr \M} \! \! \! dS \, f |_{\pr \M} ={}& \int_{\pr \M}\! \!  dS \,
\bigl ( \de t (-\pr_n f + K f) + \de_t f \bigl ) \bigl |_{\pr \M} ~. \cr }
\eqno (A.9) $$
\bigskip
\leftline{\bigbf Appendix B}
\medskip

The discussion in section 4 of the energy momentum tensor and related boundary
operators, as defined through variations of the metric for a curved space
background, seemingly requires modification if fermion fields are present.
As is well known [9] it is then necessary for a consistent treatment
to introduce
a tangent frame basis $V^\mu {}_{\! a}$ such that $g^{\mu \nu}=V^\mu {}_{\! a}
V^\nu {}_{\! a}$ and a corresponding connection $\omega_{\mu ab} = -
\omega_{\mu ba}$ defined by $\nab_\mu V^\nu {}_{\! a} = \pr_\mu V^\nu {}_{\! a}
+\Gamma_{\mu \si}^\nu V^\si {}_{\! a} + \omega_{\mu ab}V^\nu {}_{\! b}=0$. The
conventional Dirac action is
$$ S^D = \int_\M \! dv \, {\bar \psi} \bigl ( \ga^\mu {\ollr \nab}_\mu + M
\bigl ) \psi \ , \quad \gamma^\mu = V^\mu {}_{\! a}\ga_a \ , \quad
\nab_\mu = \pr_\mu + \half \omega_{\mu ab} \si_{ab} \ ,
\eqno (B.1) $$
for $\gamma_a$ the usual Dirac matrices $\{ \ga_a , \, \ga_b \} = 2 \de_{ab} ,
\ \si_{ab} = \quar [ \ga_a , \, \ga_b ]$. On the classical field equations,
$(\ga^\mu \nab_\mu + M) \psi =0 , \ {\bar \psi} (\ga^\mu {\overleftarrow \nab}
_\mu - M) =0$, then $S^D=0$.

When considering variations in $V^\mu {}_{\! a}$, which gives a variation in
the metric $g^{\mu \nu}$, it is also necessary to allow for contributions
proportional to
$$ \ell^{\mu \nu} = \half (\de V^\mu {}_{\! a} \, V^\nu {}_{\! a} -
\de V^\nu {}_{\! a} \, V^\mu {}_{\! a} ) \ .
\eqno (B.2) $$
By straightforward calculation, using $\de \omega_{\mu ab} = V_{\nu[a} \nab_\mu
\de V^\nu {}_{\! b]} - V^\si {}_{\! [a} V^\rho {}_{\! b]}\nab_\si \de
g_{\si \mu}$, assuming $\psi,\, {\bar \psi}$ satisfy the equations of motion
but without any specific boundary conditions then
$$ \eqalign{
\de_V S^D = {}& \int_\M \! dv \, \bigl ( \half \de g^{\mu \nu} T_{\mu \nu}^D
+ \half \ell^{\mu\nu} {\bar \psi}[\si_{\mu \nu}, M] \psi \bigl ) {}
- \int_{\pr \M} \! \! \! dS \, \half \ell^{ij} {\bar \psi}\ga_n
\si_{ij}\psi \ ,
\cr
& T_{\mu\nu}^D = {\bar \psi} \ga_{(\mu} {\ollr \nab}_{\nu)} \psi \ . \cr}
\eqno (B.3) $$
As usual [9], assuming invariance under local rotations which in this context
requires $[\si_{ab},\, M]=0$, terms proportional to $\ell^{\mu\nu}$ in
$\de_V S^D$ on $\M$ are absent. Clearly from (B.3), for the simple Dirac action
$S^D$, there are no contributions under a variation of the metric proportional
to $\de \hga^{ij},\dots$ as in (4.3) or (4.18).

Under variations in $\psi,{\bar\psi}$, subject to the field equations on $\M$,
$$ \de_\psi S^D =  \int_{\pr\M} \! \! \! dS \, (\de {\bar\psi} \ga_n \psi -
{\bar \psi}\ga_n \de \psi) \ .
\eqno (B.4) $$
Conventionally [10] linear boundary conditions on ${\bar \psi}',\psi$ are
chosen, as in (3.9), so that
$$ {\bar \psi}'\ga_n \psi \bigl |_{\pr \M} = 0 \ ,
\eqno (B.5) $$
and hence $i\ga^\mu\nab_\mu$ is a symmetric operator and in (B.4) we also
have $\de_\psi S^D = 0$.

The basic equations derived in section 4 from invariance under diffeomorphisms
giving expressions for $T_{ni},T_{nn}$ on $\pr \M$ in terms of boundary
operators now require extension to take account of terms proportional to
$\ell^{\mu \nu}$ as well as terms involving $\de \hga^{ij},\dots$ from
variations of the metric. Corresponding to a diffeomorphism, leading to (4.4),
we require
$$ \eqalign {
\de_v V^\mu {}_{\! a} = {}& v^\si \pr_\si V^\mu {}_{\! a} - \pr_\si v^\mu \,
V^\si {}_{\! a} = -\nab_\si v^\mu\, V^\si {}_{\! a} - v^\si\omega_{\si ab}
V^\mu {}_{\! b} \ , \cr
\de_v \psi = {}& v^\si \pr_\si \psi
= v^\si\nab_\si \psi - \half v^\si\omega_{\si ab} \si_{ab}\, \psi \ , \cr}
\eqno (B.6) $$
and hence from the definition (B.2)
$$ \ell_v^{\mu \nu} = \half(\nab^\mu v^\nu - \nab^\nu v^\mu ) +
v^\si\omega_{\si ab} V^\mu {}_{\! a} V^\nu {}_{\! b} \ .
\eqno (B.7) $$

The essential identity of section 4 has the form in this case
$$ \eqalign {
\de_v S^D = {}& - \int_{\pr \M} \! \! \! dS \, (v^n T_{nn}^D + v^i
T_{ni}^D) \cr
& + \half \int_{\pr \M} \! \! \! dS \, \bigl ( v^\mu \omega_{\mu ab}
{\bar \psi}
\half \{ \ga_n , \si_{ab} \} \psi + \hnab^i v^j \, {\bar \psi} \ga_n \si_{ij}
\psi \bigl ) \cr
& + \half \int_{\pr\M} \! \! \! dS \, (\de_v {\bar\psi} \ga_n \psi -
{\bar \psi}\ga_n \de_v \psi) \ . \cr}
\eqno (B.8) $$
This may be verified using $\de_v{\bar \psi}, \de_v\psi$ from (B.6) and
from (B.3)
$$ \eqalign {
T_{nn}^D = {}& \half {\bar \psi}\ga_n \nab_n \psi
- \half {\bar \psi}{\overleftarrow \nab}_{\! n}\ga_n \psi \ , \cr
T_{ni}^D = {}& \half {\bar \psi}\ga_n \nab_i \psi
- \half {\bar \psi}{\overleftarrow \nab}_{\! i} \ga_n \psi + \half \hnab^j
( {\bar \psi}\ga_n \si_{ij} \psi ) \ . \cr}
\eqno (B.9) $$
Since $S^D=0$ there are no terms involving $n^\mu \de S^D/\de x_\b^\mu$ as in
(4.23). To derive (B.9) it is useful to note that we may write the tangential
components of the spinor covariant derivatives on $\pr \M$ in the form
$$ \eqalign {
&\nab_i = \hnab_i + \half K_{ij} \ga_n \ga^j \ , \quad \hnab_i = \pr_i
+ {\hat \omega}_i \ , \cr
& \hnab_i \ga_n = \pr_i \ga_n + [{\hat \omega}_i , \, \ga_n ] = 0 \ , \quad
\hnab_i \ga_j = 0 \ . \cr}
\eqno (B.10) $$

If the boundary conditions on ${\bar \psi}, \, \psi$ imply also ${\bar \psi}
\ga_n \si_{ij} \psi=0$, besides (B.5), then the additional surface terms in
(B.3), (B.8) are zero and $T_{nn}^D = T_{ni}^D = 0$ on $\pr \M$ and the
discussion in section 4 does not require any modification.
\bigskip
\leftline{\bigbf Appendix C}
\medskip

Here we outline the essential steps in the calculation of the two
point function
of the energy momentum tensor at the critical point to $\rO (\vep)$. We
calculate only those contributions to $\epsilon$ and $\gamma$, as defined by
(2.8), which involve off diagonal elements of $T_{\mu \nu}$ since these are not
affected by interaction terms in this theory. By virtue of (2.10a,b) and (2.17)
knowledge of $\epsilon$ and $\gamma$ is clearly sufficient to
determine  all other
pieces of the the two point function for $T_{\mu \nu}$ as well as providing
a convenient consistency check.
\vskip 0.1in
\epsfbox{fig2.eps}

The two basic graphs whose contributions need to be calculated to $\rO (g)$
are shown in fig. 2a,b. Corresponding to fig. 2a it is sufficient to find the
one loop corrections to $\langle \phi(x) \phi(x')\rangle$. It is easy
to see that
$$ \eqalign {
\langle  \phi (y,{\bf 0}) \phi (y',{\bf 0}) \rangle^{(1)} = {}&
- {1\over 6} (n+2)g \,{1\over S_d^3 (d-2)^3}\, \bigl ( G(y,y') \pm G(y,-y')
\bigl ) \ , \cr
G(y,y') = {}& \int_{-\infty}^\infty \! \! dz \int d^{d-1}x \, {1\over X^{d-2}
X'{}^{d-2}} \, {1\over (4z^2)^{\hh (d-2)}} \ , \cr
& \quad X^2 = x^2 + (y-z)^2 \ , \quad
\quad X'{}^2 = x^2 + (y'-z)^2 \ . \cr}
\eqno (C.1) $$
Since the critical coupling $g_* = \rO(\vep)$ it is sufficient to evaluate
$G(y,y')$ for $d=4$. The integral in (C.1) then has a linear divergence which
may be regularised by imposing a cut-off $|z|>\epsilon$ and we find
$$ G(y,y') = \pi^2 \Biggl \{ {1\over (y+y')^2} \ln {4|y| |y'| \over (y-y')^2}
+ {1\over \epsilon} \, {1\over |y| + |y'|} \Biggl \} \ .
\eqno (C.2) $$
For the Dirichlet case the divergence as $\epsilon \to 0$ cancels while in
the Neumann case it may be removed by a surface counterterm in the action
$\propto \phi^2$. At the critical point to this order $g_*/16\pi^2 = 3 \vep/
(n+8)$ and the result for $\langle \phi(y,{\bf 0}) \phi(y',{\bf 0})
\rangle$ is
compatible with (2.5), since $\eta = \half(d-2) + \rO(\vep^2)$, where the
function of the conformal invariant $v$ is given by
$$ S_d (d-2) F(v)^{(1)} = - {1\over 2}\, {n+2 \over n+8} \, \vep \, \Bigl (
v^2 \ln {1-v^2\over v^2} \pm \ln (1-v^2) \Bigl ) \ .
\eqno (C.3) $$
This result is equivalent to that obtained by Gompper and Wagner [11]. For use
in calculating the contributions corresponding to fig. 2a to $\langle
T_{ij} (y,{\bf 0}) T_{k\ell} (y',{\bf 0}) \rangle$ we also require, as in (3.3)
for free fields,
$$ \eqalign {
\langle \pr_i \phi (y,{\bf 0}) \pr_j \phi (y',{\bf 0}) \rangle = {}&
- \langle  \pr_i \pr_j \phi (y,{\bf 0}) \phi (y',{\bf 0}) \rangle =
\de_{ij}\, {1\over |y-y'|^{2\eta+2}} F_1(v) \ , \cr
\langle  \pr_i \pr_j \phi (y,{\bf 0}) \pr_k \pr_\ell\phi (y',{\bf 0})
\rangle = {}&  \bigl ( \de_{ij}\de_{k\ell} + \de_{ik}
\de_{j\ell} + \de_{i\ell} \de_{jk} \bigl )\,
{1\over |y-y'|^{2\eta+4}} F_2(v) \ , \cr}
\eqno (C.4) $$
where $F_1, \, F_2$ are determined in terms of $F$ by
$$ \eqalign {
F_1(v) = {}& 2\eta F(v) - v(1-v^2) F'(v) \ , \cr
F_2(v) = {}& 2\eta (2\eta+2) F(v) - v(1-v^2) (4\eta+1 + 3v^2)F'(v) + v^2
(1-v^2) F''(v)  \ . \cr}
\eqno (C.5) $$
Corresponding to (C.3) we have, taking $\eta\to 1$,
$$ \eqalign {
S_d (d-2) F_1(v)^{(1)} = {}& -  {n+2 \over n+8} \, \vep \, \Bigl (
v^4 \ln {1-v^2\over v^2} \pm \ln (1-v^2) + v^2 \pm v^2 \Bigl ) \ , \cr
S_d (d-2) F_2(v)^{(1)} = {}& - 4\, {n+2 \over n+8} \, \vep \, \Bigl (
v^6 \ln {1-v^2\over v^2} \pm \ln (1-v^2) + v^4 + \half v^2 \pm v^2 \pm \half
v^4 \Bigl ) \ . \cr}
\eqno (C.6) $$
With these results we may find the contributions arising from fig. 2a to
$\epsilon$ and $\gamma$ just as for free fields in section 3. In the case of
$\gamma$ the required correlation functions involving $\pr_1 \phi$ may
be found from $\langle \phi (y,{\bf 0}) \phi (y',{\bf 0}) \rangle$ and
$\langle \pr_i \phi (y,{\bf 0}) \pr_j \phi (y',{\bf 0}) \rangle$ by
differentiation with respect to $y,y'$. Hence we obtain
$$ \eqalignno {
S_4^2 \epsilon_a^{(1)} = {n\over 9}\,  {n+2 \over n+8} \, \vep \, \Biggl\{&
\mp v^2(1+4v^2+v^4) \ln {(1-v^2)^2\over v^2}  \mp {9\over 2}
v^2(1+v^4) \mp 2v^4  \cr
& - 6v^8 \ln {1-v^2\over v^2} - 6 \ln (1-v^2) - 5 v^2(1+v^4) - v^4 \Biggl\}
\ , & (C.7a)\cr
S_4^2 \gamma_a^{(1)} = {n\over 9}\,  {n+2 \over n+8} \, \vep \, \Biggl\{&
\pm 5v^2(1- v^4) \Bigl ( \ln {(1-v^2)^2\over v^2} - {1\over 2} \Bigl )\cr
& - 6v^8 \ln {1-v^2\over v^2} + 6 \ln (1-v^2) + v^2(1-v^4)  \Biggl\}
 \ . & (C.7b)\cr}
$$

For the graph in fig. 2b the corresponding contributions relevant for
determining $\epsilon$ and $\gamma$ are
$$ \eqalignno{
\langle T_{ij} (y,{\bf 0}) T_{k\ell} (y',{\bf 0}) \rangle^{(1)}_b = {}& -
\bigl ( \de_{ij}\de_{k\ell} + \de_{ik} \de_{j\ell} \bigl ) \, {\ts {1\over 3}}
n(n+2) g \, \Bigl ( {d\over d-1}\Bigl )^2 {1\over d^2-1}\, {32\over S_d^4} \cr
& \times y^2 y'^2 \int_{-\infty}^\infty \! \! dz \, z^4 \int d^{d-1}x \, x^4 \,
{1\over X^{d+2} {\bar X}^{d+2}} \, {1\over X'^{d+2} {\bar X}'^{d+2}} \cr
& + \hbox {terms proportional to } \ \de_{ij} \de_{k\ell} \ , & (C.8a) \cr
\langle T_{i1} (y,{\bf 0}) T_{k1} (y',{\bf 0}) \rangle^{(1)}_b = {}& -
\de_{ik} \, {\ts {1\over 3}}
n(n+2) g \, \Bigl ( {d\over d-1}\Bigl )^2 {1\over d-1}\, {8\over S_d^4} \cr
& \times y y' \int_{-\infty}^\infty \! \! dz \, z^4 \int d^{d-1}x \, x^2 \,
{1\over X^{d+2} {\bar X}^{d+2}} \, {1\over X'^{d+2} {\bar X}'^{d+2}} \cr
&~~~~~~~~~~~~~~~~~~~~~\times (x^2 + z^2 - y^2)(x^2 + z^2 - y'^2)
\ , & (C.8b) \cr}
$$
with $X,\, X'$ as given in (C.1) and ${\bar X}^2 = x^2 + (y+z)^2 , \
{\bar X}'{}^2 = x^2 + (y'+z)^2$. It is clear from (C.8) that in the absence
of a boundary the one loop graphs for $\langle T_{\mu \nu}(x) T_{\si \rho}
(x')\rangle$ are zero [12] so there is no change in the bulk coefficient $C$
in (2.18) to this order and hence the results of all calculations for
$\epsilon, \, \gamma$ at this order should vanish as $v\to 0$. The integrals
are more tedious to evaluate in this case but there are no singularities
requiring regularisation even when $d=4$. In this case we obtain
$$ \eqalignno {
S_4^2 \epsilon_b^{(1)} = {n\over 9}\,  {n+2 \over n+8} \, \vep \, {1\over 5}
\Biggl\{& v^2(1-v^2)^2 \ln {(1-v^2)^2\over v^2}
-3v^4 + {12v^6\over (1-v^2)^2} \cr
& {} + v^2(1-v^4) \Bigl ( 1 - {2v^2\over (1-v^2)^2} + {6v^4\over (1-v^2)^4}
\Bigl ) \ln v^2  \Biggl\} \ , & (C.9a) \cr
S_4^2 \gamma_b^{(1)} = {n\over 9}\,  {n+2 \over n+8} \, \vep \, \Biggl\{&
- v^2(1-v^4) \ln {(1-v^2)^2\over v^2}
+ {v^4(1+v^2)\over (1-v^2)} \cr
& {} - v^2(1+v^4) \Bigl ( 1 - {v^2\over (1-v^2)^2} \Bigl ) \ln v^2 - v^4
\ln v^2 \Biggl\}  \ . & (C.9b) \cr}
$$
We should note that (C.7) and (C.9) are in accord with the symmetry
$\epsilon (v) = v^{2d}\epsilon(v^{-1})$, $\gamma (v) = -v^{2d}\gamma(v^{-1})$
with $d=4$. It is easy to see that the apparent singularities as $v\to 1$
cancel and in fact $\gamma_b^{(1)}=0$ for $v=1$.
\vskip 0.5cm
\leftline {\bigbf References}
\parskip 0pt
\vskip 15pt
\item{[1]}H.W. Diehl, in {\it Phase Transitions and Critical Phenomena},
vol 10, p. 75, (C. Domb and J.L. Lebowitz eds.) Academic Press,
London (1986);
\item{}H.W. Diehl, and S. Dietrich, {\it Zeit. f. Phys. B} {\bf 42}
(1981) 65, (E) {\bf 43} (1981) 281; {\bf 50} (1983) 117.
\vskip 7pt
\item{[2]} J.L. Cardy, {\it Nucl. Phys.} {\bf B240} [FS12] (1984) 514.
\vskip 7pt
\item{[3]} J.L. Cardy, {\it Phys. Rev. Lett.} {\bf 65} (1990) 1443.
\vskip 7pt
\item{[4]} E. Eisenriegler, M. Krech and S. Dietrich, preprint.
\vskip 7pt
\item{[5]} H. Osborn and A. Petkos, in preparation.
\vskip 7pt
\item{[6]} A. Capelli, D. Friedan and J.L. Latorre, {\it Nucl. Phys.}
{\bf B352} (1991) 616.
\vskip 7pt
\item{[7]} D.M. McAvity and H. Osborn, Cambridge preprint DAMTP-92/31,
{\it Nucl. Phys.}, to be published.
\vskip 7pt
\item{[8]} J.L. Cardy, seminar 1992.
\vskip 7pt
\item{[9]} S. Weinberg, {\it Gravitation and Cosmology}, John Wiley \&
Sons, New York (1972).
\vskip 7pt
\item{[10]} H. Luckock, {\it Jour. Math. Phys.}, {\bf 32} (1991) 1755.
\vskip 7pt
\item{[11]} G. Gompper and H. Wagner, {\it Zeit. f. Phys. B} {\bf 59} (1985)
193.
\vskip 7pt
\item{[12]} S.J. Hathrell, {\it Ann. Phys.} {\bf 139} (1982) 136.
\bye